\documentclass[copyright,creativecommons]{eptcs}
\providecommand{\event}{PLACES 2010}

\providecommand{\firstpage}{1}

\usepackage{breakurl}        

\usepackage[final,dvips]{graphicx}
\usepackage{amssymb}
\usepackage{amsmath}
\usepackage{listings}
\usepackage[dvipsnames]{xcolor}
\usepackage{suffix}

\newcommand{\MYPARAGRAPH}[1]{\vspace{1mm} \noindent \textbf{#1}~~}
\WithSuffix\def\MYPARAGRAPH*#1{\vspace{1mm} \noindent \textbf{#1}}

\newcommand{\REF}[1]{\S\,\ref{#1}}

\newcommand{\PARTY}[1]{\ensuremath{\textbf{#1}}}

\newif\ifrh\rhfalse

\newcommand{\RH}[1]{\ifrh{\color{red}{#1}}\else{#1}\fi}

\newcommand{\SEND}[2]{#1$\rightarrow$#2}
\newcommand{\IP}[1]{\ensuremath{\mbox{\it IP}{}_{\mbox{\tiny #1}}}}
\newcommand{\PORT}[1]{$\mbox{\it p}_{\mbox{\tiny #1}}$}
\newcommand{\ST}[2]{$\mbox{\it ST}^{\mbox{\tiny #1}}_{\mbox{\tiny #2}}$}
\newcommand{\DS}[3]{$\mbox{\it DS}^{\mbox{\tiny #1}}_{\mbox{\tiny #2}}(\mbox{\small #3})$}

\newcommand{\DSACK}[2]{${\mbox{\it ACK}}_{\mbox{\tiny #1#2}}$}

\newcommand{\LM}[2]{$\mbox{\it LM}(\mbox{\ST{#1}{#2}} - \mbox{\ST{#2}{#1}})$}

 \newcommand{\ENCan}[1]{\langle #1 \rangle}

\newcommand{\CAT}{\ensuremath{\!\cdot\!}}
\newcommand{\COL}{\ensuremath{\!:}}

\newcommand{\DUAL}[1]{\ensuremath{\overline{#1}}}

\newcommand{\EXISTS}{\ensuremath{\exists}}
\WithSuffix\def\EXISTS*#1#2{\ensuremath{\EXISTS#1.#2}}
\newcommand{\FORALL}{\ensuremath{\forall}}
\WithSuffix\def\FORALL*#1#2{\ensuremath{\FORALL#1.#2}}

\newcommand{\NEXISTS}{\ensuremath{\nexists}}
\WithSuffix\def\NEXISTS*#1#2{\ensuremath{\NEXISTS#1.#2}}

\newcommand{\SUBS}[2]{\ensuremath{\{\raisebox{0.8pt}{\ensuremath{#1}}/\raisebox{-1.4pt}{\ensuremath{#2}}\}}}
\WithSuffix\def\SUBS*#1#2{\ensuremath{\{#1/#2\}}} 

\newcommand{\VEC}[1]{\ensuremath{\vec{#1}}}

\newtheorem{DUMMY}{dummy}[section] 
\newtheorem{DEFAUX}[DUMMY]{Definition}
\newtheorem{LEMAUX}[DUMMY]{Lemma}
\newtheorem{THEAUX}[DUMMY]{Theorem}

\newcommand{\CODE}[1]{\texttt{#1}}
\newcommand{\LST}[1]{\lstinline@#1@}

\newcommand{\TAG}[1]{\textrm{#1}}
\WithSuffix\def\TAG*#1{\quad\TAG{#1}}
\newcommand{\RULE}[1]{\textsf{\small (#1)}}
\WithSuffix\def\RULE*#1{\quad\RULE{#1}}

\newcommand{\LAB}[1]{\ensuremath{#1}} 
\WithSuffix\def\LAB*#1#2{\ensuremath{\LAB{#1}_{#2}}} 

\newcommand{\VARM}[1]{\ensuremath{\textsf{#1}}} 

\newcommand{\EXCO}[1]{\ensuremath{#1}} 
\WithSuffix\def\EXCO*#1#2{\ensuremath{\EXCO{#1}[#2]}}

\newcommand{\ISEP}{\ensuremath{.}} 
\newcommand{\SSEP}{\ensuremath{;\,}} 

\newcommand{\ONEtoX}[3]{\ensuremath{#1_{1}\COL#2_{1},...,#1_{#3}\COL#2_{#3}}} 
\WithSuffix\def\ONEtoX*#1#2#3#4{\ensuremath{\TABS{#1_{1}}{#2_{1}}\COL#3_{1}, ..., \TABS{#1_{#4}}{#2_{n}}\COL#3_{#4}}} 
\newcommand{\ONEtoN}[2]{\ensuremath{\ONEtoX{#1}{#2}{n}}}
\WithSuffix\def\ONEtoN*#1#2#3{\ensuremath{\ONEtoX*{#1}{#2}{#3}{n}}}

\newcommand{\ACC}[3]{\ensuremath{#1\,(\TYPED{#2}{#3})}}
\newcommand{\REQ}[3]{\ensuremath{\DUAL{#1}\,(\TYPED{#2}{#3})}}
\newcommand{\OUT}[2]{\ensuremath{#1\,!\langle#2\rangle}}
\newcommand{\IN}[2]{\ensuremath{#1\,?(#2)}}

\newcommand{\IF}[1]{\ensuremath{\texttt{\small if}\;#1\THEN}}
\newcommand{\THEN}{\ensuremath{\;\texttt{\small then}\;}}
\newcommand{\ELSE}{\ensuremath{\;\texttt{\small else}\;}}
\newcommand{\PAR}{\ensuremath{\;|\;}}

\newcommand{\NEWS}[1]{\ensuremath{(\boldsymbol\nu\:#1)\;}}

\newcommand{\VARP}[2]{\ensuremath{#1(#2)}}
\WithSuffix\def\VARP*#1#2{\ensuremath{#1\langle#2\rangle}}
\newcommand{\NIL}{\ensuremath{\mathbf{0}}}

\newcommand{\TCASE}[2]{\ensuremath{\texttt{typecase}\;#1\;\texttt{of}\;\{#2\}}}

\newcommand{\QUEUE}[1]{\ensuremath{\vec{#1}}}
\newcommand{\TABS}[2]{\ensuremath{(#1)\:{#2}}}

\newcommand{\ACCP}[4]{\ensuremath{\ACC{#1}{#2}{#3}\ISEP#4}}
\newcommand{\REQP}[4]{\ensuremath{\REQ{#1}{#2}{#3}\ISEP#4}}
\newcommand{\OUTP}[3]{\ensuremath{\OUT{#1}{#2}\SSEP#3}}
\newcommand{\INP}[3]{\ensuremath{\IN{#1}{#2}\SSEP#3}}

\newcommand{\IFPQ}[3]{\ensuremath{\IF{#1}#2\ELSE#3}}

\newcommand{\NEWSP}[2]{\ensuremath{\NEWS{#1}#2}}

\newcommand{\TCASEXSPn}[1]{\ensuremath{\TCASE{#1}{\ONEtoN*{\VEC{\VARM{X}}}{S}{P}}}}
\WithSuffix\def\TCASEXSPn*#1{\ensuremath{\TCASE{#1}{\ONEtoN*{\VEC{\VARM{X}}}{S^{\VARM{X}}}{P}}}}

\newcommand{\TYPED}[2]{\ensuremath{#1\COL#2}}

\newcommand{\TMSG}[1]{\ensuremath{#1}} 
\WithSuffix\def\TMSG*#1#2{\ensuremath{\TMSG{#1}_{#2}}}

\newcommand{\TLAB}[1]{\ensuremath{#1}}
\WithSuffix\def\TLAB*#1#2{\ensuremath{#1_{#2}}}

\newcommand{\TCONF}[2]{\ensuremath{[#1] \, #2}}
\newcommand{\TSESS}[2]{\ensuremath{(#1, #2)}}
\WithSuffix\def\TSESS*#1#2#3{\ensuremath{(#1, \TCONF{#2}{#3})}}

\newcommand{\TSEP}{\ensuremath{;}}

\newcommand{\TOUT}[1]{\ensuremath{!\langle#1\rangle}}
\newcommand{\TIN}[1]{\ensuremath{?(#1)}}
\newcommand{\TSEL}[1]{\ensuremath{{\oplus}\{#1\}}}
\newcommand{\TBRA}[1]{\ensuremath{{\&}\{#1\}}}

\newcommand{\TEND}{\ensuremath{\texttt{end}}}

\newcommand{\TSET}[1]{\ensuremath{[#1]}}

\newcommand{\BASIS}{\ensuremath{\Theta}}

\newcommand{\HASTYPE}{\ensuremath{\triangleright}}
\newcommand{\RTENV}[3]{\ensuremath{#1\TSEP#2\TSEP#3}}
\WithSuffix\def\RTENV*{\ensuremath{\RTENV{\BASIS}{\SORTS}{\RTYPING}}}
\newcommand{\RTYPING}{\ensuremath{\Delta}}
\newcommand{\SORTS}{\ensuremath{\Gamma}}
\newcommand{\TCAT}{\ensuremath{\CAT}}
\newcommand{\TENV}[2]{\ensuremath{#1\TSEP#2}}
\WithSuffix\def\TENV*{\ensuremath{\TENV{\BASIS}{\SORTS}}}
\newcommand{\TPOSE}{\ensuremath{\,\odot\,}}
\WithSuffix\def\TPOSE*#1#2{\ensuremath{#1 \TPOSE #2}}

\newcommand{\TYPING}{\ensuremath{\Sigma}}
\WithSuffix\def\TYPING*{\ensuremath{\Delta}}

\newif\ifnma\nmafalse
\newif\ifnmaPost\nmaPostfalse

\newif\ifnmaEPTCS\nmaEPTCSfalse
\nmaEPTCStrue

\newcommand{\nma}[1]{\ifnma{\color{blue}{#1}}\else{#1}\fi}
\newcommand{\nmaPost}[1]{\ifnmaPost{\color{red}{#1}}\else{#1}\fi}

\makeatletter
\def\cleardoublepage{\clearpage\if@twoside \ifodd\c@page\else%
\hbox{}%
\thispagestyle{empty}
\newpage%
\if@twocolumn\hbox{}\newpage\fi\fi\fi}
\makeatother

\newcommand{\CLOSE}[1]{\ensuremath{\textsf{\small close}(#1)}}
\newcommand{\CLOSEP}[2]{\ensuremath{\CLOSE{#1}\SSEP#2}}
\newcommand{\MAKE}[1]{\textsf{\small make}(#1)}
\newcommand{\MAKEP}[2]{\MAKE{#1}\SSEP#2}

\newcommand{\BRASF}[3]{\ensuremath{#1\triangleright\{\textbf{Success}: #2, \textbf{Fail}: #3 \}}}

\newcommand{\SELS}[1]{\ensuremath{#1 \triangleleft \textbf{Success}}}
\newcommand{\SELSP}[2]{\ensuremath{\SELS{#1}\SSEP#2}}
\newcommand{\SELF}[1]{\ensuremath{#1 \triangleleft \textbf{Fail}}}
\newcommand{\SELFP}[2]{\ensuremath{\SELF{#1}\SSEP#2}}

\newcommand{\TBRASF}[2]{\ensuremath{\TBRA{\textbf{Success}: #1, \textbf{Fail}: #2}}}

\newcommand{\TSELSF}[2]{\ensuremath{\TSEL{\textbf{Success}: #1, \textbf{Fail}: #2}}}

\title{Secure Execution of Distributed Session Programs}

\author{Nuno Alves
\institute{Freelance Consultant}
\and
Raymond Hu
\institute{Imperial College London}
\email{rhu@doc.ic.ac.uk}
\and Nobuko Yoshida
\institute{Imperial College London}
\email{yoshida@doc.ic.ac.uk}
 \and Pierre-Malo Deni\'elou
\institute{Imperial College London}
\email{malo@doc.ic.ac.uk}
}

\def\titlerunning{Secure Execution of Distributed Session Programs}

\def\authorrunning{Alves, Hu, Yoshida and Deni\'elou}

\begin{document}
\maketitle

%
\begin{abstract}



The development of the SJ Framework for session-based distributed programming is part of recent and ongoing research into integrating \emph{session types} and practical, real-world programming languages. SJ programs featuring session types (protocols) are statically checked by the SJ compiler to verify the key property of \emph{communication safety}, meaning that parties engaged in a session only communicate messages, including higher-order communications via \emph{session delegation}, that are compatible with the message types expected by the recipient.

This \nmaPost{paper} presents current work on security aspects of the SJ Framework. Firstly, we discuss our implementation experience from improving the SJ Runtime platform with
security measures to protect and
augment
communication safety
at runtime. We implement a transport component for secure session execution that uses a modified TLS connection with authentication based on the Secure Remote Password (SRP) protocol. The key technical point is the delicate treatment of \emph{secure session delegation} to counter a previous vulnerability. We find that the modular design of the SJ Runtime, based on the notion of an Abstract Transport for session communication, supports rapid extension to utilise additional transports whilst separating this concern from the application-level session programming task.
In the second part of this abstract, we formally prove the target security properties by modelling the extended SJ delegation protocols in the $\pi$-calculus.

%
%

\end{abstract}





\section{Session Programming in SJ}
\label{sect:introduction}

It has become increasingly important to understand and specify the
behaviour of applications across many domains as sequences of
communications and interaction between concurrently executing
components, as opposed to independent black boxes that simply return a
final output from a given input. Unfortunately, the most common
technologies and programming techniques for communication-based
programming in use today do not provide the level of support and
safety guarantees enjoyed for traditional single-threaded,
``localised'' programming. For example, low-level network socket APIs
provide only the minimal mechanisms for exchanging untyped data in an
unstructured manner, and higher-level RPC/RMI libraries are typically
coupled to the synchronous call-return pattern and lack the facility
to encapsulate a series of such exchanges as one complete unit of
interaction.

SJ \cite{sj,sjhomepage} is an extension of Java that uses
\emph{session types} \cite{esop} to address the above
issues. Programmers use session types, which can be thought of as
communication protocols, to specify the abstract communication
behaviour of a program. Concrete communication behaviour is
implemented using special session primitives;
in particular,
SJ supports higher-order session types, implemented by
\emph{session delegation} actions, that express the migration of
ongoing sessions to new parties. The SJ compiler performs static
session type checking, guaranteeing that session implementations
conform to their declared types, i.e. a SJ session between compatible
peers will never reduce to a communication error other than premature
termination due to e.g. network failure.
Other works have presented SJ implementations of communication-based applications from widely
different domains, such as Internet Web services \cite{sj} and
parallel algorithms for cluster computing \cite{sessparallel}, demonstrating
competitive performance in both cases.

\MYPARAGRAPH{Securing session execution.}
To enforce the statically verified communication safety property at runtime, the SJ Runtime (SJR) validates peer compatibility at session initiation and provides a session monitoring service that dynamically tracks session progress against the expected type. Nevertheless, the development of the SJ Framework as a research project, until now, had yet to focus on \emph{session security} as a primary objective. Indeed, the security of session delegation has been a frequently posed question. In this \nmaPost{paper}, we identify a potential vulnerability in the existing SJ session delegation protocols; we then implement and formalise SJR extensions for secure session execution, solving the above problem and strengthening runtime communication safety.
\REF{sect:example} starts with an example SJ application (a modified version of the main example in \cite{sj}) that illustrates both the usual security concerns and issues specific to session delegation. \REF{sect:die} then briefly discusses the design and implementation of a new \emph{Transport Module} that enables
secure session execution
over a modified TLS connection with authentication based on the Secure Remote Password  (SRP) protocol. In \REF{sect:model}, we formalise the extended delegation protocols encapsulated by the new Transport Module, and prove the target security properties. Finally, \REF{sect:conclusion} concludes
by describing
our ongoing and future work.


\section{An Example SJ Application}
\label{sect:example}


To illustrate the session security issues, in particular regarding session delegation, we examine the interaction in a modified version of the main Web services example from \cite{sj}, an implementation of Use Case C-UC-001 from the W3C Web Services Choreography Requirements \cite{CDLUSECASE}.



\begin{figure}[t]
\centering
\includegraphics[width=0.9\textwidth]{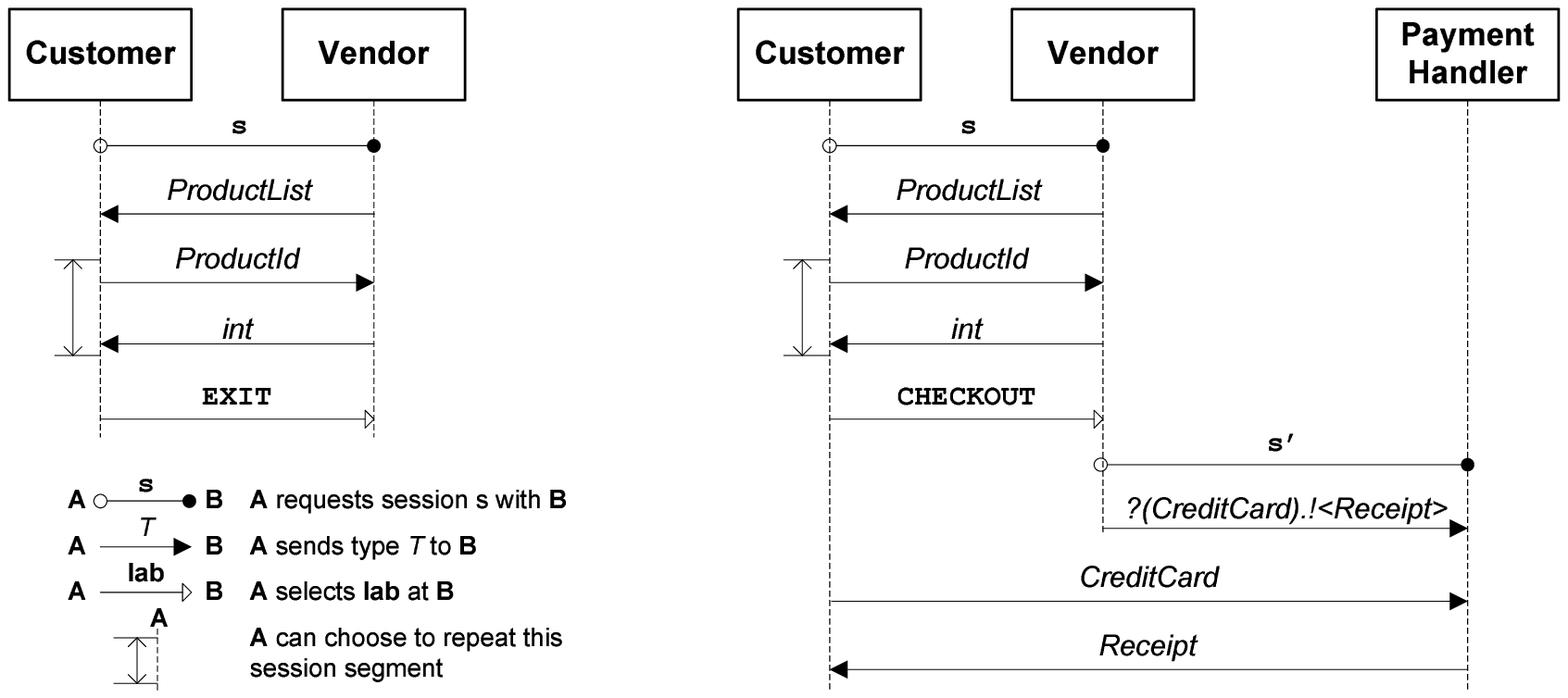}
\caption{The interaction between a Customer, the Vendor and Payment Handler service in an online purchase session.}
\label{fig:example}
\end{figure}


\MYPARAGRAPH*{The basic scenario} is an online purchase session involving three parties, a Customer (\PARTY{C}), the Vendor (\PARTY{V}) and a Payment Handler service (\PARTY{H}). The interaction between these parties constituting one session is depicted in Figure~\ref{fig:example}, where each side represents one of the two ways to complete the session. Both start by \PARTY{C} connecting to \PARTY{V}, and \PARTY{V} sending a list of the products for sale. In the next part, \PARTY{C} adds a product to the basket and \PARTY{V} returns the updated total cost of the basket; this segment can be repeated an arbitrary number of times by \PARTY{C}. After this, \PARTY{C} has two choices. On the left-hand side, \PARTY{C} cancels the purchase and ends the session by selecting the \LST{EXIT} branch. On the right-hand side, \PARTY{C} proceeds by selecting the \LST{CHECKOUT} branch. At this point, \PARTY{V} enters a session with the third party, \PARTY{H}. The single action between \PARTY{V} and \PARTY{H} is an example of \emph{session delegation}: the message type itself is a session type that specifies the remaining session actions to be completed by \PARTY{H} on behalf of \PARTY{V}. After \PARTY{V} delegates its side of the session to \PARTY{H}, the interaction proceeds between \PARTY{C} and \PARTY{H}: \PARTY{C} sends her credit card details, and \PARTY{H} issues a receipt.





\MYPARAGRAPH{Session types for this application.} Figure~\ref{fig:protocols} lists the session types, declared as SJ protocols, for the interaction between \PARTY{C} and \PARTY{V} from the perspective of each party.
The \LST{customerToVendor} protocol starts with the \LST{cbegin} element to denote the client side of the session. Receiving and sending messages have the syntax, e.g. \LST{?(ProductList)} and \LST{!<ProductId>} respectively. The repeated segment of the session is specified as a session iteration type, \LST{![...]}. Here, the \LST{!} signifies that \PARTY{C} controls the termination condition of the loop. Finally, the two session branches are collected within the \LST{!\{...\}} constructor and labelled, e.g. \LST{CHECKOUT}; the \LST{!} again signifies that \PARTY{C} makes the branch decision. The \LST{vendorToCustomer} protocol is the \emph{dual} session type that describes the reciprocal behaviour, in this case given by simply inverting the output (\LST{!}) and input (\LST{?}) symbols. Note that the \LST{?[...]} (resp. \LST{?\{...\}}) type specifies that \PARTY{V} should follow the iteration (resp. branch) decision made by \PARTY{C}.


\begin{figure}[t]
\centering
{\small
\begin{tabular}{l|l}
\begin{lstlisting}
protocol customerToVendor {
  cbegin. // Client session request.
  ?(ProductList). // Get product list.
  ![ // Can repeat this segment.
	  !<ProductId>. // Add product to basket.
	  ?(int) // Get updated basket total.
  ]*.
  !{ // Two branch options.
    CHECKOUT: // Proceed to checkout.
      !<CreditCard>.
      ?(Receipt),
    EXIT: // Cancel purchase.
  }
}
\end{lstlisting}
\qquad & \qquad
\begin{lstlisting}
protocol vendorToCustomer {
  sbegin.
  !<ProductList>.
  ?[
	  ?(ProductId).
	  !<int>
  ]*.
  ?{
    CHECKOUT:
      ?(CreditCard).
      !<Receipt>,
    EXIT:
  }
}
\end{lstlisting}
\end{tabular}
}
\caption{Session types (declared as SJ protocols) for the interaction depicted in Figure~\ref{fig:example}.}
\label{fig:protocols}
\end{figure}


Recall that \PARTY{V} will not actually perform the final \LST{?(CreditCard).!<Receipt>} exchange itself, but will delegate these actions to \PARTY{H}. These actions must still be specified in \LST{vendorToCustomer} to achieve duality between the behaviours of \PARTY{C} and \PARTY{H} (in SJ, session initiation between non-dual parties raises an exception). However, the delegation action from \PARTY{V} to \PARTY{H}, specified by the protocols
\vspace{-1mm}
\begin{center}
\begin{tabular}{ll}
\begin{lstlisting}
protocol vendorToHandler {
  cbegin.
  !<?(CreditCard).!<Receipt>>
}
\end{lstlisting}
\qquad\qquad & \qquad\qquad
\begin{lstlisting}
protocol handlerToVendor {
  sbegin.
	?(?(CreditCard).!<Receipt>)
}
\end{lstlisting}
\end{tabular}
\end{center}
\vspace{-1mm}
ensures that \PARTY{V} indeed fulfils the \LST{vendorToCustomer} protocol contract. This Use Case demonstrates how session types provide a type-safe discipline for communications programming, including higher-order communications that
evolve
the shape of the session network.

Due to space limitations, we omit the application-level implementations of these protocols in order to focus on the runtime security of delegation. The full source code for this example can be found in the \LST{tests/src} directory (see the \LST{places.purchase} package) of the SJ Google Code repository (linked from \cite{sjhomepage}); see \cite{sj} for further explanation of the SJ session primitives. \RH{The types and implementation of this application can be readily extended (e.g. to pass additional information from \PARTY{V} to \PARTY{H} before the delegation, and for \PARTY{H} to return an acknowledgment after completing the session with \PARTY{C}) by adding the required session interactions.}







\MYPARAGRAPH{A vulnerability in session delegation.} Needless to say, conducting the above session over an insecure transport connection jeopardises message confidentiality and integrity,
an especial concern for highly sensitive messages like \CODE{CreditCard}.
For this purpose, the SJR includes SSL and HTTPS variants of the basic TCP and HTTP Transport Modules, implemented using the standard Java APIs for these features.
However, the current version of SJ does not have dedicated support for \RH{mutual} peer authentication
\RH{outside of TLS certificate-based authentication (which is used to accomplish only unilateral authentication in most typical TLS authentication scenarios).}



In addition to the general attacks mentioned above, we identify a specific vulnerability in the SJ session delegation protocols. The first work on implementing session delegation \cite{sj} presented three alternative SJR protocols with varying tradeoffs for coordinating the three (or four) parties involved in the delegation of an ongoing session. Two of these protocols, the Resending Protocol and Bounded Forwarding Protocol, are termed as \emph{reconnection-based}. As the name suggests, these protocols involve closing the original transport-level connection underlying the application-level session being delegated, and establishing a new connection to reflect the session migration. A detailed recap of the delegation protocols is beyond the scope of this \nmaPost{paper}, but Figure~\ref{fig:resending} lists an instance of the Resending Protocol for the above application: $V$ (resp. $H$, $C$) denotes the SJR supporting \PARTY{V} (resp. \PARTY{H}, \PARTY{C}), \DS{$V$}{$C$}{$H$} denotes the Delegation Signal for the delegation from \PARTY{V} to \PARTY{C} with passive party \PARTY{H}, \ST{\PARTY{V}}{\PARTY{C}} the remaining type of the session between \PARTY{V} and \PARTY{C} from the former's side, \DSACK{$C$}{$V$} the Delegation Acknowledgement from $C$ to $V$, and \LM{\PARTY{C}}{\PARTY{V}} the ``lost messages'' corresponding to the difference between the two session types.
The reconnection itself is performed by the $C$ SJR over steps~6 and 7. The crucial point is that the lack of \RH{mutual authentication between all \emph{three} peers}, and hence the inability to confirm that the party accepted by $H$ (step~2) is the same as the original $C$, allows an attacker to infiltrate the session, masquerading as $C$, at this weak point. This
\nmaPost{is called Reconnection Vulnerability later in this paper and}
also applies to the Bounded Forwarding delegation protocol \cite{sj} for the same reasons.
\nmaPost{This vulnerability occurs because a normal session is designed to be binary (only two peers)
and upon delegation, a third (or fourth) party is involved.}

\RH{\MYPARAGRAPH*{For secure session delegation}, we extend the reconnection-based protocols with additional authentication message exchanges: Figure~\ref{fig:resendingSec} lists the new secure version of the original protocol instance from Figure~\ref{fig:resending}.}
\nma{Our extended protocol is different from the original in step~1, the creation of a fresh credential by $V$, and steps~2 and 5, which send the credential to $H$ and $C$ respectively. $H$ then stores the credential and waits for a connection on \PORT{$H$}. The key action in the extended protocol lies on step~9, where $C$ sends the credential to $H$ after connecting: if $H$ can validate that the credentials match, the connection is successfully established, otherwise the delegation has failed and the session is aborted.

The above protocol listings correspond to Case~1 of the Resending Protocol \cite{sj}. A complete analysis of all four Delegation Cases for our Secure Resending and Forwarding Protocols is specified in \cite{nmaMscThesis}.}

\begin{figure}[t]
\centering	
\begin{tabular}{@{\extracolsep{1mm}}rrlc|crrl}
1. & \SEND{$V$}{$H$}: & \multicolumn{6}{l}{\LST{START\_DELEGATION}} \\
2. & $H$: & \multicolumn{6}{l}{Open server socket on free port \PORT{$H$}; accept connection on \PORT{$H$}} \\
3. & \SEND{$H$}{$V$}: & \multicolumn{6}{l}{\PORT{$H$}} \\
4. & \SEND{$V$}{$C$}: & \multicolumn{6}{l}{\DS{$V$}{$C$}{$H$} $=$ $\ENCan{\mbox{\ST{\PARTY{V}}{\PARTY{C}}}, \mbox{\IP{$H$}}, \mbox{\PORT{$H$}}}$} \\		
5. & \SEND{$C$}{$V$}: & \DSACK{$C$}{$V$} \\
6. & $C$: & Close $s$
& & &
6'. & $V$: & Close $s$\\
7. & $C$: & Connect to \IP{$H$}\CODE{:}\PORT{$H$} & \\
8. & \SEND{$C$}{$H$}: & \LM{\PARTY{C}}{\PARTY{V}} & \\
\end{tabular}
\caption{Operation of the original Resending Protocol between the Vendor,
Handler and Customer.\label{fig:resending}}
\end{figure}

\begin{figure}[t]
\centering	
\begin{tabular}{@{\extracolsep{1mm}}rrlc|crrl}
1. & $V$: & Credential creation \\
2. & \SEND{$V$}{$H$}: & \multicolumn{6}{l}{\LST{START\_DELEGATION::CRED}} \\
3. & $H$: & \multicolumn{6}{l}{Open server socket on free port \PORT{$H$}; accept connection on \PORT{$H$}} \\
4. & \SEND{$H$}{$V$}: & \multicolumn{6}{l}{\PORT{$H$}} \\
5. & \SEND{$V$}{$C$}: & \multicolumn{6}{l}{\DS{$V$}{$C$}{$H$} $=$ $\ENCan{\mbox{\ST{\PARTY{V}}{\PARTY{C}}}, \mbox{\IP{$H$}}, \mbox{\PORT{$H$}}, \mbox{\LST{CRED}}}$} \\
6. & \SEND{$C$}{$V$}: & \DSACK{$C$}{$V$} \\
7. & $C$: & Close $s$
& & &
7'. & $V$: & Close $s$\\
8. & $C$: & Connect to \IP{$H$}\CODE{:}\PORT{$H$} & \\
9. & \SEND{$C$}{$H$}: & \LST{CRED}
& & &
9'. & $H$: & \LST{CRED} checking\\
9a. & -pass: & Connection successful
& & &
9a'. & -pass: & Connection established\\
9b. & -fail: & Credential rejected, close $s$
& & &
9b'. & -fail: & Authentication error, close \PORT{$H$}\\
10. & \SEND{$C$}{$H$}: & \LM{\PARTY{C}}{\PARTY{V}} & \\
\end{tabular}
\caption{Operation of the \emph{Secure} Resending Protocol between the Vendor,
Handler and Customer.\label{fig:resendingSec}}
\end{figure}


\section{Implementation of a Secure Transport Module}
\label{sect:die}


We first summarise the SJ Framework and the structure of the SJ Runtime (SJR). We then briefly explain the design and implementation of a secure Transport Module plugin for the SJR that solves the security issues described in \REF{sect:example}.

\MYPARAGRAPH*{The SJ Framework} comprises the compiler and the SJ Runtime (SJR). The compiler transforms
SJ programs into a \emph{transport-independent} form in standard Java, translating the application-level SJ session primitives in terms of Java control flow and calls to \emph{Interaction Services} hosted by the SJR, also implemented in Java. Thus, SJ programs can be executed on any standard JVM, where the purpose of the SJR is to perform the requested Interaction Services as actions on an underlying transport connection. The key element in the SJR is the \emph{Abstract Transport Interface}, which represents an idealised asynchronous, reliable and order-preserving message-oriented transport for session communication. Interaction Service components are implemented as actions on the Abstract Transport, which are in turn implemented by \emph{Transport Modules} that encapsulate the communication mechanisms of specific ``concrete'' transports, such as TCP and shared memory. The Abstract Transport thus serves to decouple the realisation of session interaction semantics in the SJR from the provision of the underlying communication mechanisms.

\MYPARAGRAPH{Design.} Our primary goal was to provide a SJR Transport Module that incorporates a means of peer authentication in addition to message confidentiality and integrity, and thereby solve the security concern exposed by the reconnection-based delegation protocols.
After evaluating the available options (see \cite{nmaMscThesis} for the omitted details), our chosen design adapts
TLS
to use the Secure Remote Password (SRP) \cite{srp} authentication method \cite{rfc5054}.
Although standard usage of TLS provides confidentiality and integrity, standard
authentication
relies on certificates produced by external authorities.  We choose to
\nma{combine TLS and SRP in order to provide session security without requiring
  trusted third-parties, whilst still being able to support the
  certificate-based authentication in our implementation. 
}
\RH{Retaining session independence from external third-parties is also in line
  with the established session theory (which our formalism in \REF{sect:model}
  follows), where scope restriction of each session to just the two parties
  involved is an important invariant in proving communication safety
  \cite{tlca}.} 





\MYPARAGRAPH{Implementation and the extended delegation protocols.} TLS consists of three basic phases: negotiation of the algorithms supported, key exchange and authentication, and symmetric cipher encryption/message authentication. Unfortunately, none of the publically available Java implementations of SSL/TLS fulfilled all of our requirements: pure Java implementations either lacked support for SRP integration or were otherwise incomplete, and using JNI to interface e.g. C libraries would break SJ portability \cite{nmaMscThesis}.
As a consequence, we decided to implement SRP as an initial mutual authentication phase before the standard TLS phases. At the client side, we use a modified SJ transport negotiation protocol to activate the SRP; at the server side, we override the behaviour of the \CODE{\small accept} method of the standard Java \LST{SSLServerSocket} API.

With the SRP authentication
in place, the resending-based delegation protocols are strengthened by using the (successfully authenticated) session-sender (i.e. the party delegating the session --- in our example, \PARTY{V}) to generate and distribute fresh, secure credentials on behalf of the party performing the reconnection (\PARTY{C}), as illustrated in \REF{sect:example}. By presenting these credentials to the session-receiver (\PARTY{H}) over the \emph{new} TLS connection (ruling out replay attacks), the latter can confirm the authenticity of the connecting party, \nmaPost{which was a gap in the authentication of the final two peers after the session delegation}. The detailed protocol specifications for the extended delegation protocols can be found in \cite{nmaMscThesis}.

\section{A Process Model for Secure Sessions}
\label{sect:model}

\vspace{-1mm}


To prove the design of the new Transport Module satisfies the intended security properties, we model the extended delegation protocols in the $\pi$-calculus and formalise each property. The three original properties for correct session delegation are Linearity, Liveness and Session Consistency, which were proved by case analysis for each of the four valid
delegation scenarios (referred to as \emph{Delegation Cases 1--4}) \cite{sj}. To these, we add a new \emph{Session Security} property to rule out attacks on the delegation protocols: there are three aspects to this property, corresponding to the key
security mechanisms that the delegation protocols depend on. In the following, we focus on
\nma{the three}
aspects of the latter;
the other properties (plus the cases omitted from below) are fully documented in \cite{nmaMscThesis}.
%
%
%
See a formalisation of the Resending Protocol 
in the appendix.
As a convention, we use $A$, $B$ and $C$ to
respectively denote the passive party, session-sender and
session-receiver in the three-party delegation scenarios (Delegation
Cases 1--3 in \cite{nmaMscThesis}).
\nma{(In the previous example of \REF{sect:example}, the passive party, session-sender and session-receiver are respectively the Customer, the Vendor and the Payment Handler; we now use a general notation for the delegation roles rather than the parties from that specific example.)}
Our approach has been influenced by \cite{esop, tlca}. Below, we use ``$\forall \text{protocols}$'' to mean
all of the protocols formalised in \cite{nmaMscThesis}.

\MYPARAGRAPH{Freshness} \emph{(The credentials are fresh for only the current session.)}

\vspace{-1mm}
\begin{center}
$\forall \text{protocols} . \exists P \equiv \MAKE{Cred};Q \text{ s.t. $P$ is a session-sender}$
\end{center}
\vspace{-1mm}
where $\MAKE{Cred}; Q$ is a binder for $Cred$ in $Q$ whose reduction
instantiates $Cred$ to a fresh value in $Q$. Freshness means that
every session-sender $P$ (defined as $P \equiv
\OUTP{s}{Cred}{P_1\text{; } \OUTP{s'}{...,Cred,...}{P_2}})$) in the
protocols creates a new credential for that specific
session. Freshness in the Resending Protocol is verified for
Delegation Case~1 as follows (we omit the parallel composed processes):
\vspace{-1mm}
\begin{center}
$B = \MAKEP{CredA}{\OUTP{s'_\text{BC}}{CredA}{B_1}} \longrightarrow \OUTP{s'_\text{BC}}{CredA}{B_1} \longrightarrow $ \\
$B_1 \longrightarrow^{*} \OUTP{\DUAL{s_\text{AB}}}{S',\IP{C},x_{p_{C}},CredA}{B_2} \longrightarrow B_2$
\end{center}
\vspace{-1mm}
$B$ creates a new credential specific to the current session before the actual delegation action, which it sends to both $C$ and $A$ (for $A$, it is sent together with other important session information). So, we can conclude that Freshness holds in this case since the credential is fresh by the semantics of \textsf{\small make}.

\MYPARAGRAPH{\nmaPost{Ternary Authentication for Delegation} } \emph{(Session delegation only succeeds if the credentials of the passive party matches that of the session-receiver. Otherwise, a delegation error has occurred and the session is terminated.)}
\vspace{-1mm}
\begin{center}
\begin{math}
\begin{array}{ll}
& \forall \text{protocols} . \exists P_C \equiv \INP{s}{x_{Cred}}{P;} \ACCP{port}{x}{\DUAL{S}}{\INP{x}{y_{Cred}}{ \\
& \qquad\qquad\qquad\qquad\qquad
    \IFPQ{x_{Cred}==y_{Cred}}
    { x \triangleleft \textbf{Success}; Q }
    { x \triangleleft \textbf{Fail}; \CLOSE{x}}}} \\
\text{and } & \exists P_A \equiv \REQP{port}{x}{S}{\OUTP{x}{y_{Cred}}{ x \triangleright \{ \textbf{Success}: Q, \textbf{Fail}: \CLOSE{x} \} } } \\
\text{s.t.} & \text{$P_C$ is the session-receiver and $P_A$ is the passive party} \\
\end{array}
\end{math}
\end{center}
\vspace{-1mm}
where $P$ is defined to be a session-receiver if $P \equiv \INP{s}{Cred}{P_1\text{; } \INP{s'}{Cred}{P_2}}$, and a passive party if $P \equiv \INP{s'}{...,Cred,...}{P_1\text{; } \OUTP{s}{Cred}{P_2}}$. This property states that in all protocols, the session-receiver must start by receiving a set of credentials from the session-sender. The session-receiver then accepts a new connection on the open $port$, and receives another set of credentials. If the credentials match, then the session is continued by selecting \textbf{Success} branch; otherwise the new connection is closed, aborting the session. The passive party starts by requesting a connection on $port$ followed by sending the credential it received from the session-sender. If the credential is accepted it similarly follows the \textbf{Success} branch; otherwise it goes to the \textbf{Fail} branch, which closes the new connection from the other end.

Below we verify Credential Checking for the Resending Protocol
Delegation Case~2 in \cite{nmaMscThesis}.
We omit $B$'s specification and interactions,
and focus on the interactions between
$A$ and $C$. 
%
\vspace{-1mm}
\begin{center}
\setlength{\tabcolsep}{1mm}ohh

\noindent \begin{tabular}{llcl}
Let & $P$ & $=$ & \BRASF{x}{\OUT{x}{LM(S-S')}}{\CLOSE{x}} \\
and & $Q$ & $=$ & \IFPQ{x_{CredA}==y_{CredA}}{\SELSP{x}{\IN{x}{x_\text{LM}}}}{\SELFP{x}{\CLOSEP{x}{\CLOSE{p_{C}}}}}  \\
\end{tabular}

\vspace{1mm}

\begin{math}
\begin{array}{lll}
A \ | \ B \  | \INP{\DUAL{s'_{BC}}}{x_{CredA}}{C_1} & \longrightarrow^{*} &
\REQP{p_{C}}{x}{S}{\OUTP{x}{z_{CredA}}{P}} \ | \ B' \ | \
\ACCP{p_{C}}{x}{\DUAL{S}}{ \INP{x}{y_{CredA}}{Q}}\\
& \longrightarrow^{*} &
\OUTP{x}{z_{CredA}}{P} \ | \ B' \ | \ \INP{x}{y_{CredA}}{Q}\\
& \longrightarrow^{*} &  (P  \ | \ B' \ | \ Q) \longrightarrow^{*} \NIL \\
\end{array}
\end{math}
\end{center}
\vspace{-1mm}
The delegation protocol starts with $C=\INP{\DUAL{s'_{BC}}}{x_{CredA}}{C_1}$ receiving a credential from $B$. Then $A$ connects to $C$ via $p_C$ and establishes a new session (bound to $x$). To finalise the delegation, $A$ sends its credentials to $C$ who compares this value with the one received from $B$: if they match, the delegation is successful; otherwise the session is closed along with the port $p_C$.

\MYPARAGRAPH{\nmaPost{Reconnection Vulnerability} } \emph{(Protection against attacks from the Network, ensuring message authentication, confidentiality and integrity.)}

The formalism assures protection of sessions from Network attacks due to the firm restriction that each session channel can only be accessed by the processes involved in that session, i.e. each channel is private to those parties and cannot be interfered by others.
To fully prove this property, an additional layer is needed to model the lower level protocols underneath the session calculus; one suitable approach would be the Applied $\pi$-Calculus \cite{applied}, which we leave as future work. As a simple demonstration, however, we can model a Network Attacker $E$ that attempts to interfere with the genuine session parties.
%
\vspace{-1mm}
\begin{center}
E = \INP{s_{ABC}}{S', x_{IP_C}, y_{p_C}, z_{CredA}}{\OUTP{s_{ABC}}{ACK}{\REQP{y_{P_C}}{z}{S}{\OUTP{z}{z_{CredA}}{}}}}  \\
\qquad\qquad\BRASF{z}{\OUT{z}{LM(S-S')}}{\NIL}
\end{center}
\vspace{-1mm}
If $E$ is able to intercept the delegation information, then she will be able to masquerade as $A$ without $C$ knowing. Note that $E$ does not close the session after the $ACK$ as the attack may benefit from remaining connected to all parties.
%
However, since there is no interaction with any external $s_{ABC}$ in any of the delegation protocols, we can infer that $E$ will remain blocked forever.
This is in line with the design choice to keep
SJ close to the original session types theory by restricting the scope of all sessions from foreign parties.




\section{Conclusion}
\label{sect:conclusion}






The aim of this work was to examine and improve session security in SJ. 
In addition to standard security concerns, we identified a vulnerability in the resending-based Resending and Bounded Forwarding protocols by modelling and analysing the delegation protocols in the $\pi$-calculus. To overcome this problem, we implemented a new Transport Module for secure session execution that combines TSL with SRP authentication. We found that the structure of the SJ Framework
cleanly decouples
application-level logic related to implementing specific sessions
from
the provision of general, lower-level communication mechanisms. Hence, we were able to readily extend the SJR so that all existing SJ programs can immediately utilise this new Transport Module without any modifications to the application source code or other SJR components. To verify the correctness of our approach, we incorporated our new extensions into the formal model to formalise and prove the intended security properties.


\MYPARAGRAPH{Future Work.}
We are currently looking at modifying a TLS implementation
to include the SRP protocol in the cipher suite list. This way, the
additional key exchange could be avoided and SRP would be included in
the key exchange phase of the TLS protocol.
\nmaPost{We are also investigating several other ways of securing the delegation
protocols instead of using the TLS-SRP. 
}
Timestamping and revocation of credentials can also be explored.

Another direction for our work is to handle the securisation of the extension of
SJ to multiparty session types~\cite{honda08multiparty}. Previous work has been
done~\cite{CDFBL07,crypto} that secures multiparty session execution in F\#
despite compromised participants. We plan to adapt it to the SJ framework and
extend it to support delegation.


\nmaPost{We finally want to model the delegation protocols at a lower level in
  order to prove the absence of the Reconnection Vulnerability (protection
  against network attacks, ensuring message integrity, authentication and
  confidentiality). This is not possible by using the standard
  $\pi$-calculus since it already assumes sessions to be isolated from each others.
  We plan to use the applied $\pi$-calculus~\cite{applied}, in which we can
  model primitives for exchanging keys during the cryptographic protocol and
  detail all other steps of the protocols.}

\label{sect:bib}
\bibliographystyle{eptcs}
\bibliography{secure_sj}



\appendix
\section{Appendix}
\label{sect:appendix}

\subsection{Resending Protocol - Case 1}

This is the most used case in which there is a simple delegation between three parties and the passive party is waiting for a receiving instruction.\\

\noindent \textbf{A} - \TSET{\TSET{ P \PAR s_\text{AB} : \langle S \rangle \QUEUE{h} } } \\

A = \INP{s_\text{AB}}{S',x_\text{IPc},y_\text{pc},z_\text{CredA}} {
    \OUTP{s_\text{AB}}{ACK}{
    \CLOSEP{s_\text{AB}} {
    \REQP{y_\text{pc}}{x}{S}{
    \OUTP{x}{z_\text{CredA}} {
    \\ \BRASF{x}{\OUT{x}{LM(S-S')}}{\CLOSE{x}}
    } } } } } \\

\subsubsection{Typing}

\begin{math}
  \SORTS \vdash \NIL \HASTYPE \TYPED{s_\text{AB}} \TEND \TCAT \TYPED{x} \TEND \\[0.2cm]
  \SORTS \vdash \BRASF{x}{\OUT{x}{LM(S-S')}}{\CLOSE{x}}
  \\ \hspace*{1.0cm} \HASTYPE \TYPED{s_\text{AB}} \TEND \TCAT \TYPED{x} \TBRASF{\TOUT{String}}{\TEND} \\[0.2cm]
  \SORTS \vdash \OUTP{x}{z_\text{CredA}} { \BRASF{x}{\OUT{x}{LM(S-S')}}{\CLOSE{x}} } \\
  \hspace*{1.0cm} \HASTYPE \TYPED{s_\text{AB}} \TEND \TCAT \TYPED{x} \TOUT{CredT} \TSEP \TBRASF{\TOUT{String}}{\TEND} \\[0.2cm]
  \SORTS \vdash \REQP{y_\text{pc}}{x}{S}{ \OUTP{x}{z_\text{CredA}} { \BRASF{x}{\OUT{x}{LM(S-S')}}{\CLOSE{x}} } }
  \HASTYPE \TYPED{s_\text{AB}} \TEND \\[0.2cm]
  \SORTS \vdash \OUTP{s_\text{AB}}{ACK}{ \CLOSEP{s_\text{AB}}{ \REQP{y_\text{pc}}{x}{S}{ \OUTP{x}{z_\text{CredA}} \\ \hspace*{1.0cm} { \BRASF{x}{\OUT{x}{LM(S-S')}}{\CLOSE{x}} } } } }
  \HASTYPE \TYPED{s_\text{AB}} \TOUT{ACK} \TSEP \TEND \\[0.2cm]
  \SORTS \vdash \INP{s_\text{AB}}{S',x_\text{IPc},y_\text{pc},z_\text{CredA}} { \OUTP{s_\text{AB}}{ACK}{ \CLOSEP{s_\text{AB}}{ \REQP{y_\text{pc}}{x}{S}{ \OUTP{x}{z_\text{CredA}} \\ \hspace*{1.0cm} { \BRASF{x}{\OUT{x}{LM(S-S')}}{\CLOSE{x}} } } } } }
  \HASTYPE \TYPED{s_\text{AB}} \TIN{S,nat,nat,CredT} \TSEP \TOUT{ACK} \TSEP \TEND \\
\end{math}

\noindent This process starts by receiving the information regarding the session receiver, along with the credential, followed by sending an ACK confirming the reception. Then, it closes the session it has with B, followed a connection to the receiver and sending the credential. If the connection is accepted, it sends information about the lost messages; otherwise the new connection is closed.

\noindent \textbf{B} - \TSET{\TSET{ Q \PAR \DUAL{s_\text{AB}} : \langle S' \rangle \QUEUE{h'} \PAR s'_\text{BC} : \langle S'' \rangle \QUEUE{h''} } } \\

B = \MAKEP{CredA}{
    \OUTP{s'_\text{BC}}{CredA}{
    \ACCP{b}{x_\text{pc}}{S}{
    \OUTP{\DUAL{s_\text{AB}}}{S',IPc,x_\text{pc},CredA}
    \INP{\DUAL{s_\text{AB}}}{x_\text{ACK}}{
    \CLOSE{\DUAL{s_\text{AB}}}
    } } } } \\

\subsubsection{Typing}

\begin{math}
  \SORTS \vdash \NIL \HASTYPE \TYPED{\DUAL{s_\text{AB}}} \TEND \TCAT \TYPED{s'_\text{BC}} \TEND \TCAT \TYPED{x_\text{pc}} \TEND \\[0.2cm]
  \SORTS \vdash \INP{\DUAL{s_\text{AB}}}{x_\text{ACK}}{ \CLOSE{\DUAL{s_\text{AB}}} }
  \HASTYPE \TYPED{\DUAL{s_\text{AB}}} \TIN{ACK} \TSEP \TEND \TCAT \TYPED{s'_\text{BC}} \TEND \TCAT \TYPED{x_\text{pc}} \TEND \\[0.2cm]
  \SORTS \vdash \OUTP{\DUAL{s_\text{AB}}}{S',IPc,x_\text{pc},CredA}{ \INP{\DUAL{s_\text{AB}}}{x_\text{ACK}}{ \CLOSE{\DUAL{s_\text{AB}}} } }
  \\ \hspace*{1.0cm} \HASTYPE \TYPED{\DUAL{s_\text{AB}}} \TOUT{S,nat,nat,CredT} \TSEP \TIN{ACK} \TSEP \TEND \TCAT \TYPED{s'_\text{BC}} \TEND \TCAT \TYPED{x_\text{pc}} \TEND \\[0.2cm]
  \SORTS \vdash \ACCP{b}{x_\text{pc}}{S}{ \OUTP{\DUAL{s_\text{AB}}}{S',IPc,x_\text{pc},CredA}{ \INP{\DUAL{s_\text{AB}}}{x_\text{ACK}}{ \CLOSE{\DUAL{s_\text{AB}}} } } }
  \\ \hspace*{1.0cm} \HASTYPE \TYPED{\DUAL{s_\text{AB}}} \TOUT{S,nat,nat,CredT} \TSEP \TIN{ACK} \TSEP \TEND \TCAT \TYPED{s'_\text{BC}} \TEND \\[0.2cm]
  \SORTS \vdash \OUTP{s'_\text{BC}}{CredA}{ \ACCP{b}{x_\text{pc}}{S}{ \OUTP{\DUAL{s_\text{AB}}}{S',IPc,x_\text{pc},CredA}{ \INP{\DUAL{s_\text{AB}}}{x_\text{ACK}}{ \CLOSE{\DUAL{s_\text{AB}}} } } } }
  \\ \hspace*{1.0cm} \HASTYPE \TYPED{\DUAL{s_\text{AB}}} \TOUT{S,nat,nat,CredT} \TSEP \TIN{ACK} \TSEP \TEND \TCAT \TYPED{s'_\text{BC}} \TOUT{CredT} \TSEP \TEND \\
\end{math}

\noindent B starts by creating a fresh credential for that delegation, which is sent to C and later to A along with other important information on how to connect to C. In the end it receives an ACK from A indicating that everything went well, so it can close the session it has with A.

\noindent \textbf{C} - \TSET{\TSET{ R \PAR \DUAL{s'_\text{BC}} : \langle S^3 \rangle \QUEUE{h^3} } } \\

C = \INP{\DUAL{s'_\text{BC}}}{x_\text{CredA}} {
    \NEWSP{pc}{(
    \REQP{b}{pc}{S}{
    \ACCP{pc}{x}{S}{
    \INP{x}{y_\text{CredA}} { \\
    \IFPQ{x_\text{CredA}==y_\text{CredA}}
        { \SELSP{x}{\IN{x}{x_\text{LM}}} }
        { \SELFP{x}{\CLOSEP{x}{\CLOSE{pc}}} }
    } } } } }) \\

\subsubsection{Typing}

\begin{math}
  \SORTS \vdash \NIL \HASTYPE \TYPED{\DUAL{s'_\text{BC}}} \TEND \TCAT \TYPED{pc} \TEND \TCAT \TYPED{x} \TEND \\[0.2cm]
  \SORTS \vdash \IN{x}{x_\text{LM}}
  \HASTYPE \TYPED{\DUAL{s'_\text{BC}}} \TEND \TCAT \TYPED{pc} \TEND \TCAT \TYPED{x} \TIN{String} \TSEP \TEND \\[0.2cm]
  \SORTS \vdash \SELSP{x}{\IN{x}{x_\text{LM}}}
  \HASTYPE \TYPED{\DUAL{s'_\text{BC}}} \TEND \TCAT \TYPED{pc} \TEND \TCAT \TYPED{x} \TSELSF{\TIN{String} \TSEP \TEND}{\TEND} \\[0.4cm]
  \text{Let } Q = \IFPQ{x_\text{CredA}==y_\text{CredA}}
        { \SELSP{x}{\IN{x}{x_\text{LM}}} }
        { \SELFP{x}{\CLOSEP{x}{\CLOSE{pc}}} } \\[0.2cm]
  \SORTS \vdash Q
  \HASTYPE \TYPED{\DUAL{s'_\text{BC}}} \TEND \TCAT \TYPED{pc} \TEND \TCAT \TYPED{x} \TSELSF{\TIN{String} \TSEP \TEND}{\TEND} \\[0.2cm]
  \SORTS \vdash \INP{x}{y_\text{CredA}} { Q }
  \HASTYPE \TYPED{\DUAL{s'_\text{BC}}} \TEND \TCAT \TYPED{pc} \TEND \TCAT \TYPED{x} \TIN{CredT} \TSEP \TSELSF{\TIN{String} \TSEP \TEND}{\TEND} \\[0.2cm]
  \SORTS \vdash \ACCP{pc}{x}{S}{ \INP{x}{y_\text{CredA}} { Q } }
  \HASTYPE \TYPED{\DUAL{s'_\text{BC}}} \TEND \TCAT \TYPED{pc} \TEND \\[0.2cm]
  \SORTS \vdash \REQP{b}{pc}{S}{ \ACCP{pc}{x}{S}{ \INP{x}{y_\text{CredA}} { Q } } }
  \HASTYPE \TYPED{\DUAL{s'_\text{BC}}} \TEND \\[0.2cm]
  \SORTS \vdash \INP{\DUAL{s'_\text{BC}}}{x_\text{CredA}} { \NEWSP{pc}{(\REQP{b}{pc}{S}{ \ACCP{pc}{x}{S}{ \INP{x}{y_\text{CredA}} { Q) } } } } }
  \HASTYPE \TYPED{\DUAL{s'_\text{BC}}} \TIN{CredT} \TSEP \TEND \\
\end{math}

\noindent The session receiver starts by receiving the credential from the session sender. Then, it creates a new port $pc$ to accept a new connection from the passive party. It then sends that port to B on the shared channel $b$, followed by the waiting of the connection from A. This process terminates by receiving the credential from A and checking if this credential matches with the one received from B: if it does, it accepts the information regarding the lost messages; otherwise the new session is closed, together with the opened port $pc$.



\end{document}